\documentclass[fleqn,12pt,twoside]{article}
\usepackage{espcrc1}

\usepackage{graphicx}
\usepackage[figuresright]{rotating}

\title{
Anti-$d$ and Anti-$He$ Production in $Au$+$Au$ Collisions at RHIC
}

\author{
Jens S\"oren Lange, Christof~Struck, on behalf of the STAR Collaboration  \\
\quad \\ 
Institut f\"ur Kernphysik\thanks{Currently on sabattical leave to 
Brookhaven National Laboratory, Upton, New York 11973, USA}, 
Johann Wolfgang Goethe-Universit\"at,\\
August-Euler-Stra\ss{}e 6, 
60486 Frankfurt/Main, Germany
}
       
\begin{document}

\maketitle

\noindent
Presented at the 
8$^{th}$ International Conference on Clustering Aspects of\\
Nuclear Structure and Dynamics, November 28, 2003, Nara, Japan.

\begin{abstract}
\noindent
Ultra-relativistic $Au$+$Au$ collisions at RHIC 
($\sqrt{s}$=130, 200 GeV) are used 
to study production of rare anti-nuclei. 
These clusters of anti-nucleons are formed by coalescence, 
i.e. overlapping wave functions of anti-nucleons. 
The coalescence coefficients $B_2$ for Anti-$d$ and $B_3$
for Anti-$^3He$ are determined, and used to derive
the fireball radius.  
\end{abstract}

%%%%%%%%%%%%%%%%%%%%%%
\section{Introduction}
%%%%%%%%%%%%%%%%%%%%%%

\noindent
The STAR experiment at the Relativistic Heavy Ion Collider (RHIC) 
at Brookhaven National Laboratory, New York, USA, 
investigates novel QCD phenomena
at high density and high temperature
in ultra-relativistic $Au$+$Au$ collisions.
RHIC has a diameter of 3.8~km, and in total
1740 superconducting magnets \cite{rhic_overview}. 
The highest center-of-mass energy 
($\sqrt{s}$=200~GeV) is about a factor 10 higher than past
fixed target experiments (e.g.\ CERN SPS $E_{beam}$=160~GeV,
equivalent to $\sqrt{s}$$\simeq$18~GeV). 
The main STAR subdetector is a midrapidity ($|$$\eta$$|$$\leq$1.6) Time Projection 
Chamber (TPC, $R$=2~m, $L$=4~m) \cite{tpc} with $\simeq$48,000,000 pixels. 
In 3 years of data taking, high statistics (10$^6$$\leq$$N$$\leq$10$^7$ 
events on tape) for Au+Au, p+p and d+Au at $\sqrt{s}$=200 GeV
and Au+Au at $\sqrt{s}$=130 GeV were recorded,
which enabled analyses of rare events.

\noindent
The particle identification of anti-particles is performed by energy loss
$dE/dx$, corresponding to charged particle ionization in the TPC gas 
(10\% methane, 90\% argon).
As production of Anti-$He$ is a rare process ($dN/dy$$\sim$10$^{-6}$),
a level-3 trigger system \cite{l3} was used for realtime 
anti-nuclei identification.
One central $Au$+$Au$ collision contains about $\simeq$130,000 TPC hits
and $\simeq$6,500 TPC tracks, which were fully reconstructed in realtime 
($t$$\leq$100~ms) by 432 Intel i960 CPUs and 48 ALPHA 21264 CPUs, respectively. 
The trigger signature was a tag onto a negative charge $Q$=-2, 
by usage of the TPC $dE/dx$ in realtime.
As no other particles in nature are expected to have the same signature,
the signature is clean.  
For a single track, $\leq$45 $dE/dx$ samples (corresponding to 45 TPC
hits) were used for the calculation of a 70\% truncated mean 
(i.e.\ excluding samples in the Landau tail of the $dE/dx$ distribution).
A track quality cut of $N_{Hit}$$\geq$23 hits per track on the level-3 trigger
and $N_{Hit}$$\geq$30 in the offline analysis was applied.
Anti-$He$ candidate tracks were also required to have a 
distance-of-closest-approach
to the event collision vertex of $\leq$7.5~cm (trigger) and $\leq$3.0~cm
(offline). 
The TPC $dE/dx$ resolution was 11\% (trigger) and 8\% (offline).
The particle identification cuts were 
0.7$<$$dE/dx$/$I_{BB}$ (trigger) and
0.88$<$$dE/dx$/$I_{BB}$$<$1.28 (offline), where $I_{BB}$ denotes
the expected Bethe-Bloch ionization.
For a particle or anti-particle of $p_T$=1~GeV/c, 
the level-3 trigger $p_T$ resolution
is $\simeq$2\% (at an solenoidal magnetic field of $B$=0.5~T) and
the track finder efficiency $\varepsilon$$\simeq$75\%. 
At the beginning of high luminosity RHIC spills, an level-3 trigger 
Anti-$He$ candidate enhancement factor of $\sim$4.2 was achieved.

\vspace*{-0.95cm}
\begin{figure}[hhh]
\begin{minipage}[b]{80mm}
\noindent
%%%%%%%%%%%%%%%%%%%%%%
\section{Anti-Nuclei}
%%%%%%%%%%%%%%%%%%%%%%
\label{ccoalescence}
\noindent
As nuclei are completely made of matter,
there is trivially no anti-matter in the initial state.
The dominant anti-nuclei production mechanism is 2-step, namely
{\it (1)} pair production of $p$$\overline{p}$, $n$$\overline{n}$, 
followed by {\it (2)} coalescence, i.e.\
the anti-nucleon wave functions overlap inside a homogenity volume.
For the systematic study of the $A$ dependence of production yields,
it is useful to define an {\it invariant yield}

\begin{equation}
E \frac{d^3 N_A}{d^3 P} = B_A
( E \frac{d^3 N_N}{d^3 p} )^A  
\end{equation}

with $p$=$P$/$A$.
\end{minipage}
\hspace{\fill}
\begin{minipage}[b]{75mm}
%\framebox[79mm]{\rule[-26mm]{0mm}{52mm}}
\framebox[75mm]{\includegraphics[scale=0.58]{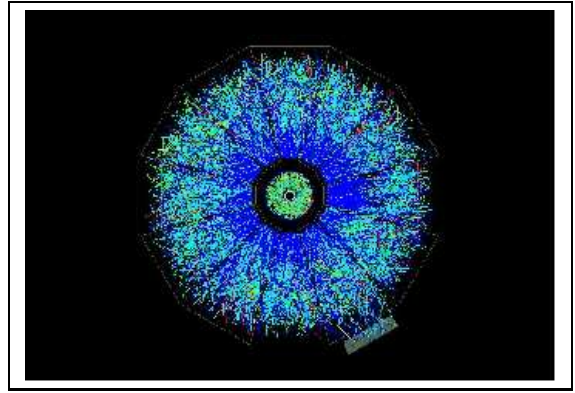}}
\caption{
A RHIC $Au$+$Au$ Collision at $\sqrt{s}$=200~GeV.
Track reconstruction was performed by the STAR Level-3 trigger system 
within $\leq$100 ms.
}
\label{fevent}
\end{minipage}
\end{figure}

\begin{figure}[hhh]
\begin{minipage}[t]{75mm}
%\framebox[75mm]{\rule[-26mm]{0mm}{52mm}}
%\framebox[75mm][c]{\centerline{\includegraphics[scale=0.405]{dedx_he3.ps}}}
\centerline{\includegraphics[width=80mm,height=55mm]{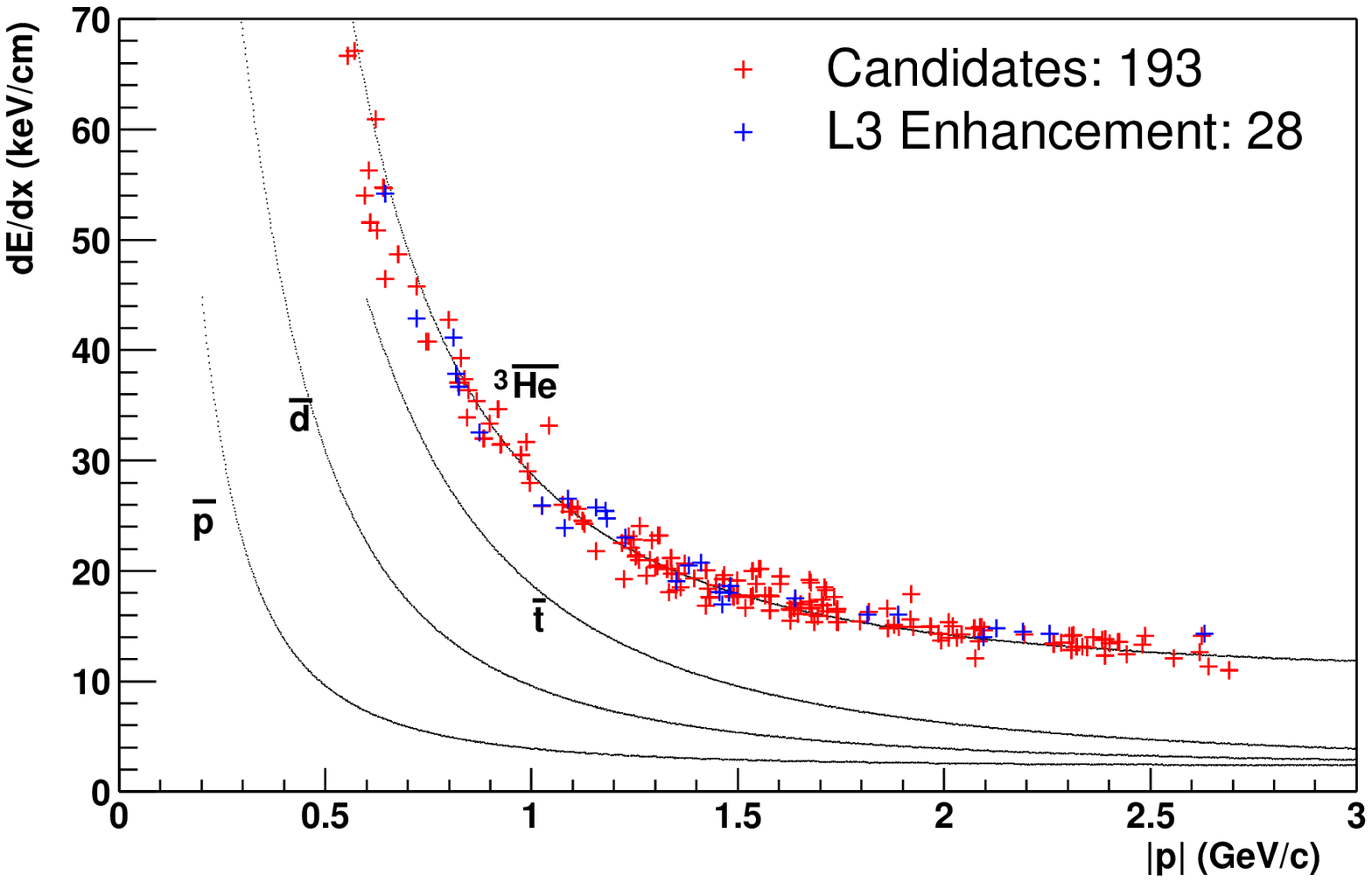}}
\caption{Ionization $dE/dx$ vs.\ momentum $p$ for the final Anti-$He$ yield.}
\label{fdedx_he3}
\end{minipage}
\hspace{\fill}
\begin{minipage}[t]{80mm}
\centerline{\includegraphics[width=80mm,height=55mm]{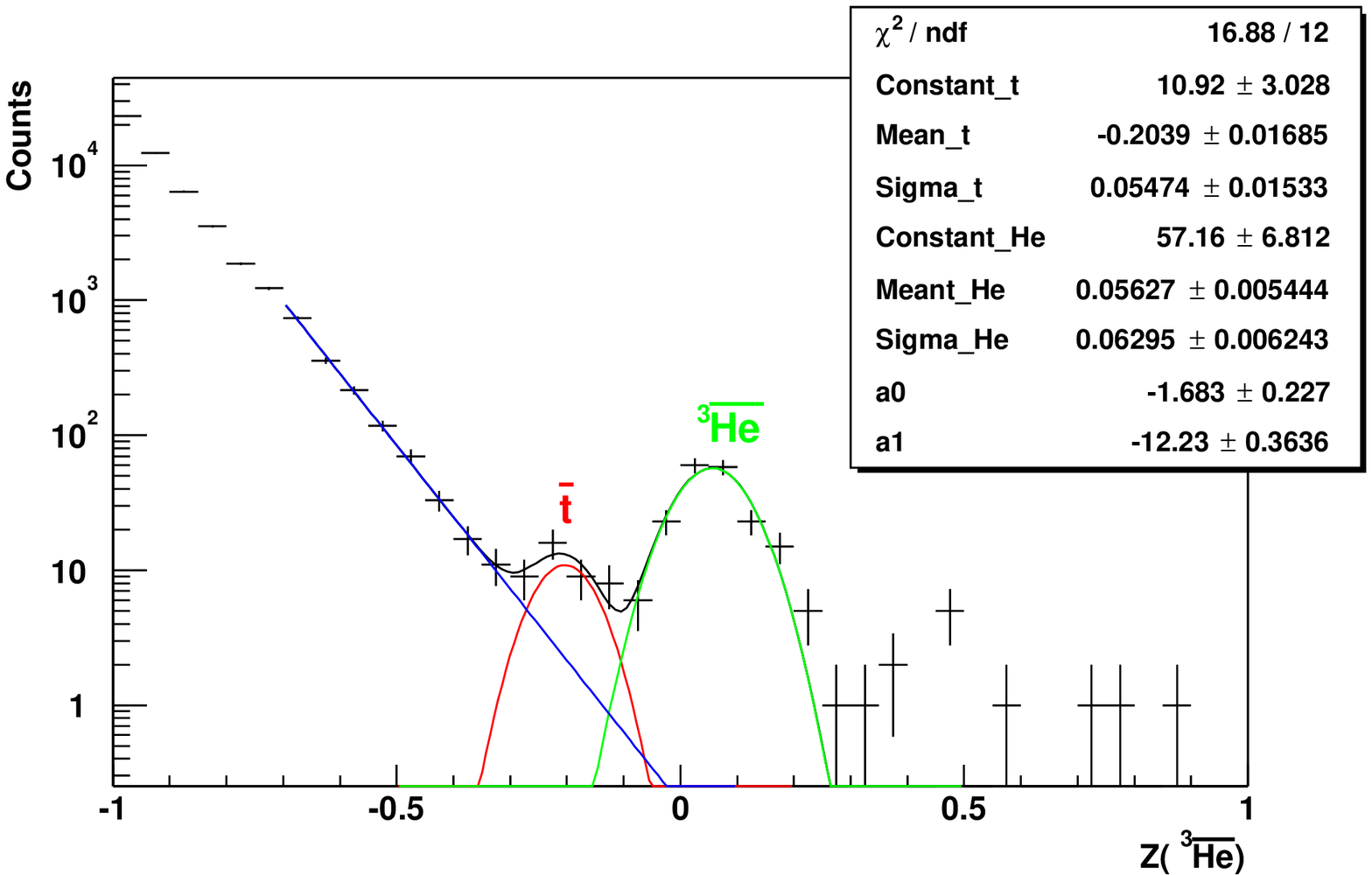}}
\caption{Distribution of $Z$=ln($dE/dx$/$I_{BB}$) (ratio of measured and
expected ionization) for the final Anti-$^3He$ yield.
Contamination from Anti-$t$ is visible, and taken into account in 
the systematic error in Tab.~2.}
\label{fyield_He}
\end{minipage}
\end{figure}

\vspace*{-0.85cm}
\noindent
The coalescence coefficient $B_A$ denotes the probability, that $A$ anti-nucleons form a bound state.
It might be regarded as a ``penalty factor'' for the step from $n$ to $n+1$ anti-nucleons
in the system. Typical orders of magnitude are $B_2$$\sim$10$^{-3}$ and $B_3$$\sim$10$^{-6}$.
For high $\sqrt{s}$ and a large system size, the coalescence coefficient is related 
to the inverse of the effective volume containing the anti-nuclei (i.e.\ the fireball
volume) by $B_A$=1/$V_{eff}^{A-1}$.
Thus, the coalescence coefficient can be used to determine the size of the
system (Ch.~\ref{csize}), and compared with sizes from other, complementary methods such as
$\pi^{\pm}\pi^{\pm}$ interferometry \cite{hbt}. 

\noindent
An additional motivation for the investigation of Anti-$He$ production
in the laboratory is the possibility of comparison to
Anti-$He$ production in the universe.
Recently, a search for Anti-$He$ in the earth orbit was performed 
by the space shuttle-bound experiment AMS \cite{ams}.
Two possible production mechanisms are known.
On the one hand, as $\sim$24\% of all $He$ in the universe is primordial 
and was formed at $t$$\simeq$1~s.
Thus, an interesting question is, if primordial Anti-$He$ exists, too.
On the other hand, $p$$A$, $\mu$$A$, $A$$A$ collisions\footnote{Heavy nuclei 
in the stellar 
medium are mainly produced in supernova explosions.} in the stellar medium 
permanently create anti-nuclei.
Measurements of the coalescence parameters $B_A$ at high $\sqrt{s}$ 
are needed for estimates of abundances for both production mechanisms \cite{astro}.

\noindent
Results for Anti-$d$ and Anti-$He$ production at $\sqrt{s}$=130~GeV 
are published in \cite{anti-130}. Preliminary results for 
$\sqrt{s}$=200~GeV are reported in \cite{cris}.
This paper reports the final results at $\sqrt{s}$=200~GeV
based upon a factor 2 increased statistics \cite{dr_struck}, 
using a final data set of:\\
$\bullet$ at $\sqrt{s}$=130~GeV (year 2000), 0.63$\cdot$10$^6$ 
$Au$+$Au$ collisions without trigger 
(solenoidal field $B$=0.25~T, 18\% most central events), and\\ 
$\bullet$ at $\sqrt{s}$=200~GeV (year 2001), 
3.36$\cdot$10$^6$ $Au$+$Au$ events without trigger, and 
0.72$\cdot$10$^6$ events with trigger
($B$=0.5~T, 10\% most central events).\\
In 0.72$\cdot$10$^6$ events with trigger, 
867 Anti-$He$ canditate events were triggered 
with a trigger efficiency of $\varepsilon$$\simeq$80\%.
For Anti-$d$, no trigger was used,
as the required trigger signature of charge $Q$=-1 
would be contaminated by $\pi^-$, $K^-$ and anti-protons.

%%%%%%%%%%%%%%%%%%%%%%%%
\section{Anti-$d$ Yield}
%%%%%%%%%%%%%%%%%%%%%%%%

\noindent
Anti-$d$ could only be identified cleanly 
in a small range 0.4$\leq$$p_T$$\leq$1.0~GeV/c, 
as $dE/dx$ bands merge for $p_T$$\geq$1.0~GeV/c.
An additional rapidity cut if $|$$y$$|$$\leq$0.3 
was required in order to filter long tracks.
The uncorrected raw yield is 
$N_{{\rm Anti-}d}$=6416.
Invariant yields are listed in Tab.~\ref{tyield_d},
in comparison with published yields at $\sqrt{s}$=130~GeV \cite{anti-130}.
An absorption correction, based upon a parametrized annihilation
cross section, is applied \cite{dr_struck}.

\begin{table}
\begin{minipage}[b]{80mm}
\begin{tabular}{|l|l|l|}
\hline
$\sqrt{s}$ & $p_T$ & $E$$\frac{d^3 N}{d^2 p}$ \\
(GeV) & (GeV/c) & (GeV$^{-2}$c$^3$) \\
\hline
\hline
130 & 0.55 & (2.47$\pm$0.26)$\times$10$^{-3}$ \\
130 & 0.65 & (1.87$\pm$0.19)$\times$10$^{-3}$ \\
130 & 0.75 & (1.90$\pm$0.20)$\times$10$^{-3}$ \\
\hline
\hline
200 & 0.55 & (3.2$\pm$0.3)$\times$10$^{-3}$ \\
200 & 0.65 & (3.2$\pm$0.3)$\times$10$^{-3}$ \\
200 & 0.75 & (3.0$\pm$0.4)$\times$10$^{-3}$ \\
200 & 0.85 & (2.5$\pm$1.0)$\times$10$^{-3}$  \\
\hline
\end{tabular}
\label{tyield_d}
\caption{Invariant yield for Anti-$d$.}
\end{minipage}
\hspace{\fill}
\begin{minipage}[b]{75mm}
\begin{tabular}{|l|l|l|}
\hline
$\sqrt{s}$ & $p_T$ & $E$$\frac{d^3 N}{d^2 p}$ \\
(GeV) & (GeV/c) & (GeV$^{-2}$c$^3$) \\
\hline
\hline
130 & & (8.4$\pm$2.3)$\times$10$^{-7}$ \\
\hline
\hline
200 & 1.5 & (1.9$\pm$0.3)$\times$10$^{-6}$ \\
200 & 2.5 & (1.0$\pm$0.2)$\times$10$^{-6}$ \\
200 & 3.5 & (0.47$\pm$0.07)$\times$10$^{-6}$ \\
200 & 4.5 & (0.09$\pm$0.03)$\times$10$^{-6}$ \\
\hline
\end{tabular}
\label{tyield_He}
\caption{Invariant yield for Anti-$He$.}
\end{minipage}
\end{table}

%%%%%%%%%%%%%%%%%%%%%%%%%%%%
\section{Anti-$^3He$ Yield.}
%%%%%%%%%%%%%%%%%%%%%%%%%%%%

\noindent
For Anti-$^3He$, a clean $dE/dx$ identification  
is possible over the wide range 1.0$\leq$$p_T$$\leq$5.0~GeV/c.
Thus, a wider rapidity cut $|$$y$$|$$\leq$1.0 
could be applied without compromising the signal significance.
The final raw yield is $N_{{\rm Anti-}^3He}$=193
and shown in Fig.~2 and Fig.~3. 
Invariant yields are listed in Tab.~\ref{tyield_He},
in comparison with published yields at $\sqrt{s}$=130~GeV \cite{anti-130}.

%%%%%%%%%%%%%%%%%%%%%%%%%%%%%%%%%%%%%%%%%%%%%
\section{Coalescence Parameters $B_2$, $B_3$}
%%%%%%%%%%%%%%%%%%%%%%%%%%%%%%%%%%%%%%%%%%%%%

\noindent
Using a measured invariant yield for anti-protons  
$E$$\frac{d^3 N}{d^2 p}$ = 0.7-2.5 GeV$^{-2}$c$^3$ \cite{dr_struck} 
as additional input, the coalescence parameters 
$B_2$ for Anti-$d$ (Fig.~4) and 
$B_3$ for Anti-$^3He$ (Fig.~5)
could be determined from the yields. 
In can be seen, that RHIC represents the highest $\sqrt{s}$,
a factor 10 higher than past experiments.
Both $B_2$ and $B_3$ decrease as a function of $\sqrt{s}$.
According to $B_A$$\sim$$1/V^{A-1}$ (Ch.~\ref{ccoalescence}),
this can be interpreted by increasing fireball volume.

\begin{table}[hhh]
\begin{minipage}[b]{80mm}
%%%%%%%%%%%%%%%%%%%%%%%%%%%%%%
\section{Volume of Homogenity}
%%%%%%%%%%%%%%%%%%%%%%%%%%%%%%
\label{csize}
\noindent
In a model, which combines coalescence, hydrodynamics
and source expansion \cite{scheibl_heinz}, the $A$ dependence
of the coalescence mechanism can be calculated.
Thus, a homogenity volume $V_{hom}$ can be defined,
which does not depend on $A$ anymore.
\begin{equation}
V_{eff} ( A, M_T) = ( \frac{2\pi}{A} )^{3/2} V_{hom} ( m_T )
\label{evhom}
\end{equation}
\end{minipage}
\hspace{\fill}
\begin{minipage}[b]{75mm}
\begin{tabular}{|l|l|l|l|}
\hline
$\sqrt{s}$ & $A$ & $V_{hom}$ & $R$=$\sqrt[3]{V_{hom}}$ \\
(GeV) & & (fm$^3$) & (fm) \\
\hline
130 & $B_2$ & 107$\pm$7 & 4.75$\pm$0.10 \\
130 & $B_3$ & 99$\pm$16 & 4.63$\pm$0.25 \\
200 & $B_2$ & 124$\pm$50 & 4.99$\pm$0.67 \\
200 & $B_3$ & 121$\pm$40 & 4.95$\pm$0.54 \\
\hline
\end{tabular}
\label{tvhom}
\caption{Homogenity volumes and radii.}
\end{minipage}
\end{table}

\vspace*{-0.4cm}
\noindent
The parameter $m_T$=$\sqrt{p_T^2 + m^2}$ 
denotes the transverse mass of the anti-nucleon, 
$M_T$ the transverse mass of the anti-nucleus.
Tab.~3 lists the homogenity volumes and radii $R$=$\sqrt[3]{V_{hom}}$,
which were derived from the coalescence coefficients by using
$B_A$=1/$V_{eff}^{A-1}$ and Eq.~\ref{evhom}.
The radii can be compared to radii derived from 
$\pi^{\pm}$$\pi^{\pm}$ interferometry \cite{hbt}, i.e.\
in beam direction 
$R_{long}$=5.99$\pm$0.19(stat.)$\pm$0.36(syst.),
and perpendicular to the beam direction 
$R_{out}$=5.39$\pm$0.18(stat.)$\pm$0.28(syst.).
The radii in Tab.~3 are smaller, which can be explained
by the expectation that the radii relate to the transverse mass
by $R$$\sim$1/$\sqrt{m_T}$ \cite{scheibl_heinz}. 
The $\pi$ mesons from $\pi^{\pm}$$\pi^{\pm}$ interferometry   
correspond to 0.2$\leq$$m_T$$\leq$0.5 GeV, 
the anti-nuclei to $m_T$$\geq$1~GeV.\\

\newpage

\noindent
In summary, ultra-relativistic $Au$+$Au$ collisions at RHIC were used 
to study production of anti-nuclei. 
A raw yield of 6416 Anti-$d$ and 193 Anti-$^3He$ was collected.
Coalescence coefficients $B_2$ and $B_3$ were determined 
at a $\sqrt{s}$ which is a factor 10 higher than other 
experiments, and might be used as input for estimates
of anti-nuclei production in the stellar medium.
$B_2$ and $B_3$ were used to derive fireball radii of $\simeq$4.6-5.0~fm, which are 
consistent with radii from $\pi^{\pm}$$\pi^{\pm}$ interferometry $\simeq$5.4-6.0~fm,
when taking into account different transverse mass $m_T$ of pions and anti-nuclei.\\ 

\vspace*{-0.3cm}
\begin{figure}[ttt]
\begin{minipage}[t]{80mm}
\centerline{\includegraphics[width=80mm,height=55mm]{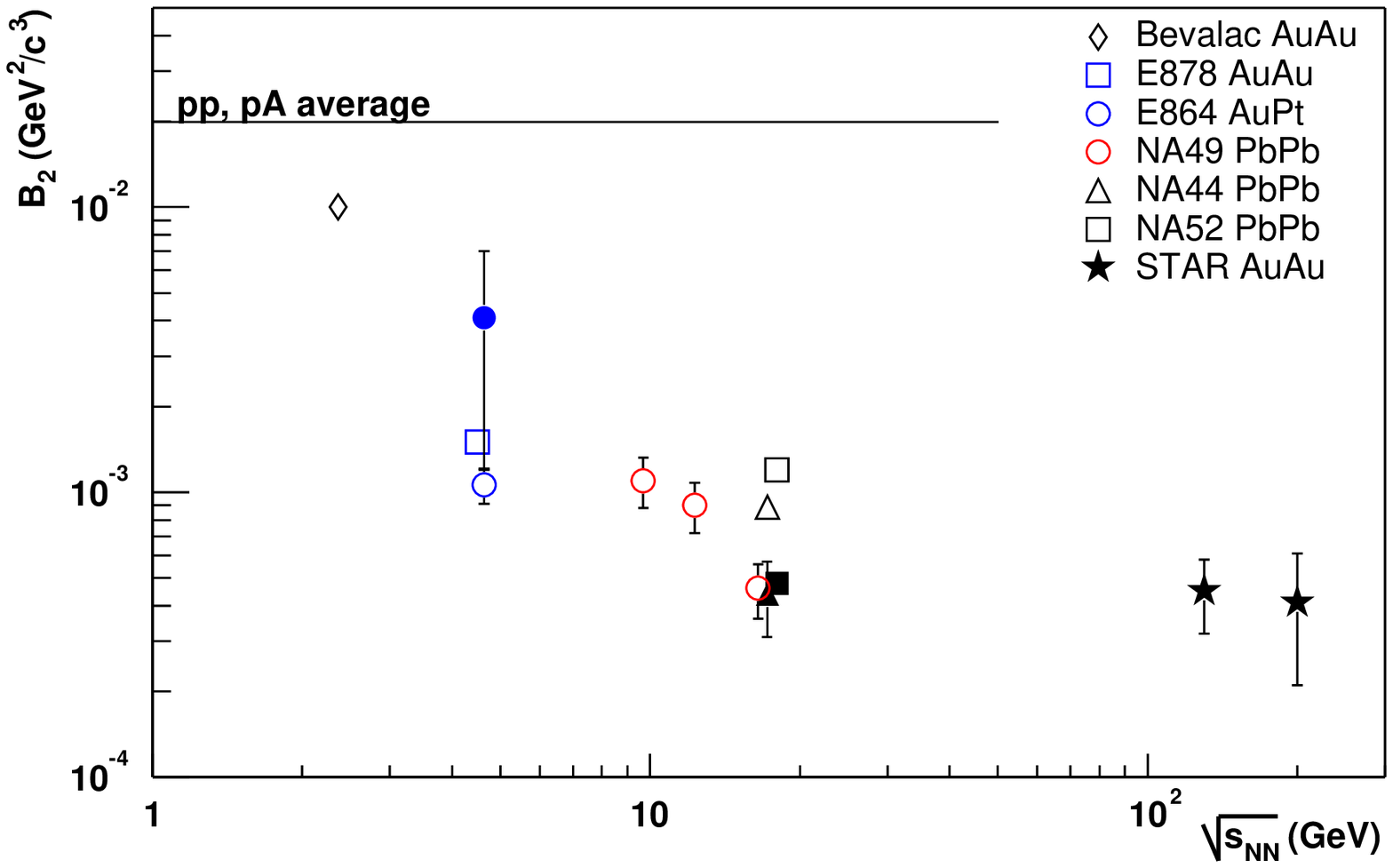}}
\label{fb2}
\caption{Coalescence coefficient $B_2$ for Anti-$d$ 
production vs.\ $\sqrt{s}$ for different experiments.}
\end{minipage}
\hspace{\fill}
\begin{minipage}[t]{75mm}
\centerline{\includegraphics[width=80mm,height=55mm]{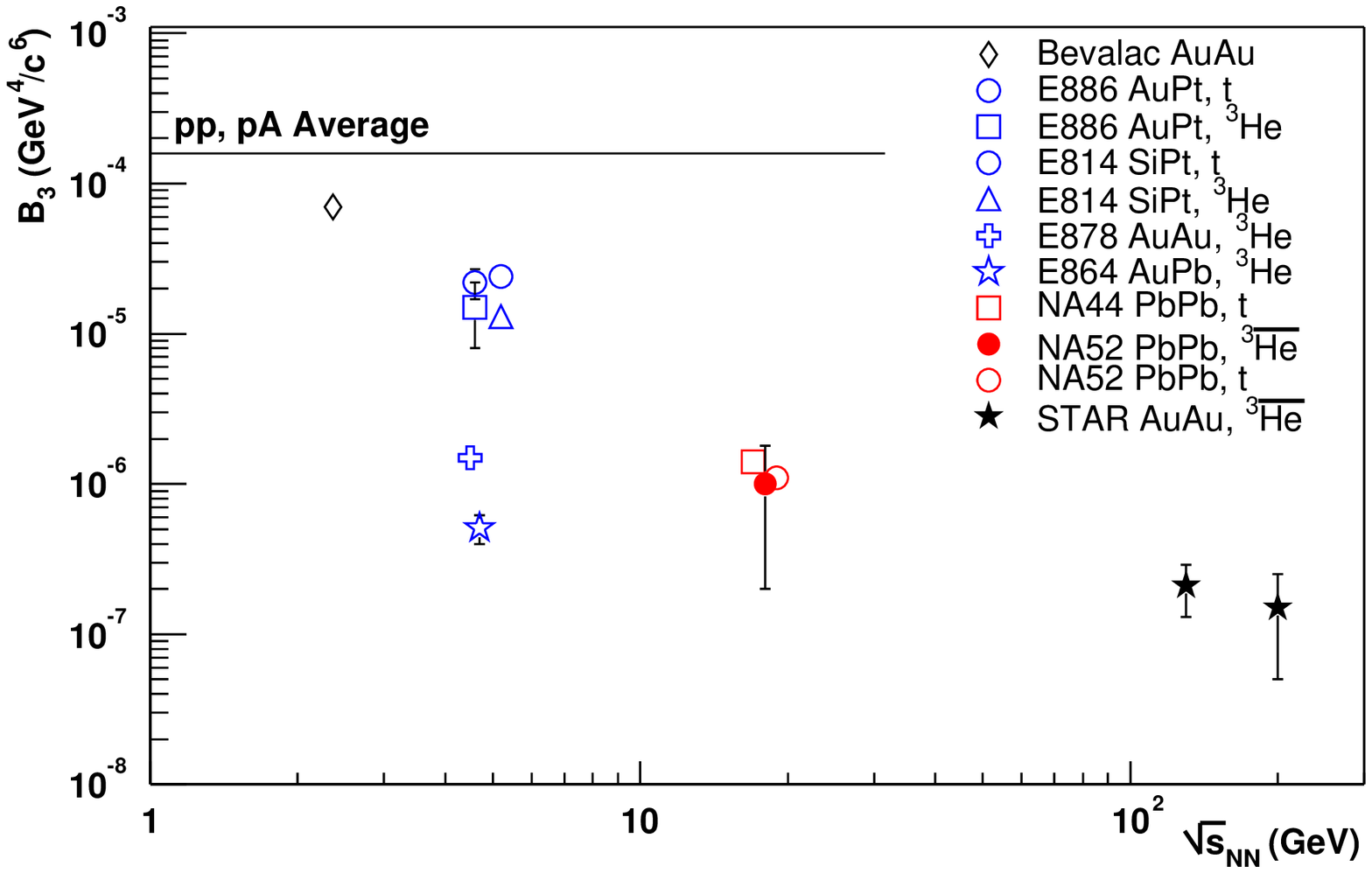}}
\label{fb3}
\caption{Coalescence coefficient $B_3$ for Anti-$^3He$ and Anti-$t$
production vs.\ $\sqrt{s}$ for different experiments.}
\end{minipage}
\end{figure}

\noindent
As a future outlook, STAR will resume data taking in 12/2003. 
In an envisaged data sample of 50,000,000 $Au$+$Au$ Events, 
one would expect an yield of $\geq$3 Anti-$^4He$, 
which would be a first observation.

%\begin{slide} 
%\begin{minipage}{25cm}
%\begin{color}{red}
%{\bf ...}
%\vspace*{0.2cm}
%\hrule
%\end{color}
%\begin{minipage}{19cm}
%{\small \quad}\\
%\begin{small}
%$\bullet$ {\bf ...}\\
%\h ... \\
%\h ... \\
%\h ... \\
%$\bullet$ ... \\
%\end{small}
%\end{minipage}
%\begin{minipage}{7cm}
%\vspace*{0.5cm}
%\raisebox{0cm}{\resizebox{7cm}{!}{\includegraphics{file.ps}}}
%\end{minipage}
%\end{minipage}
%\vfill\hrule 
%\sfoot
%\end{slide}

%\begin{table}[htb]
%\caption{Caption.}
%\label{table:1}
%\newcommand{\m}{\hphantom{$-$}}
%\newcommand{\cc}[1]{\multicolumn{1}{c}{#1}}
%\renewcommand{\tabcolsep}{2pc} % enlarge column spacing
%\renewcommand{\arraystretch}{1.2} % enlarge line spacing
%\begin{tabular}{@{}|l|l|l|}
%\hline
%1 & 2 & 3 \\
%\hline 
%\end{tabular}\vspace*{0.2mm}\\
%\end{table}

%\begin{figure}[htb]
%\begin{minipage}[t]{80mm}
%\framebox[79mm]{\rule[-26mm]{0mm}{52mm}}
%\includegraphics[width=15pc]{file}|\\
%\includegraphics[height=5pc]{file}|\\
%\includegraphics[scale=0.6]{file}|\\
%\includegraphics[angle=90,width=20pc]{file}|
%\caption{Caption.}
%\label{fig:largenenough}
%\end{minipage}
%\end{figure}

\end{document}